\documentclass[%
reprint,
superscriptaddress,
nobalancelastpage,
 amsmath,amssymb,
aps,
pra,
]{revtex4-2}

\usepackage{graphicx}
\usepackage{dcolumn}
\usepackage{bm}
\usepackage{amsfonts}
\usepackage{graphicx}
\usepackage{dcolumn}
\usepackage{bm}
\usepackage{lineno}
\usepackage{color}

\begin{document}

\preprint{APS/123-QED}

\title{\textcolor{black}{Quantum state discrimination in a $\mathcal{PT}$-symmetric system}}

\author{Dong-Xu Chen}
 \email{xdc.81@stu.xjtu.edu.cn}
 \affiliation{Quantum Information Research Center, Shangrao Normal University, Shangrao, Jiangxi 334001, China}
\author{Yu Zhang}
\affiliation{School of Physics, Nanjing University, Nanjing, Jiangsu 210093, China}
\author{Jun-Long Zhao}%
\affiliation{Quantum Information Research Center, Shangrao Normal University, Shangrao, Jiangxi 334001, China}
\author{Qi-Cheng Wu}%
\affiliation{Quantum Information Research Center, Shangrao Normal University, Shangrao, Jiangxi 334001, China}
\author{Yu-Liang Fang}%
\affiliation{Quantum Information Research Center, Shangrao Normal University, Shangrao, Jiangxi 334001, China}
\author{Chui-Ping Yang}%
 \email{yangcp@hznu.edu.cn}
 \affiliation{Quantum Information Research Center, Shangrao Normal University, Shangrao, Jiangxi 334001, China}
 \affiliation{Department of Physics, Hangzhou Normal University, Hangzhou, Zhejiang 311121, China}
\author{Franco Nori}%
 \email{fnori@riken.jp}
 \affiliation{Theoretical Quantum Physics Laboratory, RIKEN, Wako-shi, Saitama 351-0198, Japan}
  \affiliation{RIKEN Center for Quantum Computing (RQC), Wako-shi, Saitama 351-0198, Japan}
   \affiliation{Physics Department, The University of Michigan, Ann Arbor, Michigan 48109-1040, USA}

\date{\today}

\begin{abstract}
Nonorthogonal quantum state discrimination (QSD) plays an important role in quantum information and quantum communication. In addition, compared to Hermitian quantum systems, parity-time-($\mathcal{PT}$-)symmetric non-Hermitian quantum systems exhibit novel phenomena and have attracted considerable attention. Here, we experimentally demonstrate QSD in a $\mathcal{PT}$-symmetric system (i.e., $\mathcal{PT}$-symmetric QSD), by having quantum states evolve under a $\mathcal{PT}$-symmetric Hamiltonian in a \textcolor{black}{lossy} linear optical setup. We observe that two initially nonorthogonal states can rapidly evolve into orthogonal states, and the required evolution time can even be vanishing provided the matrix elements of the Hamiltonian become sufficiently large. \textcolor{black}{We also observe that the cost of such a discrimination is a dissipation of quantum states into the environment.} Furthermore, by comparing $\mathcal{PT}$-symmetric QSD with optimal strategies in Hermitian systems, we find that at the critical value, $\mathcal{PT}$-symmetric QSD is equivalent to the optimal unambiguous state discrimination in Hermitian systems. \textcolor{black}{We also extend the $\mathcal{PT}$-symmetric QSD to the case of discriminating three nonorthogonal states.} \textcolor{black}{The QSD} in a $\mathcal{PT}$-symmetric system opens a new door for quantum state discrimination, which has important applications in quantum computing, quantum cryptography, and quantum communication.
\end{abstract}

\maketitle


\section{Introduction}
Quantum state discrimination (QSD) \cite{barnett2009quantum} is a central issue in quantum mechanics. Its applications cover quantum computing, quantum cryptography, and quantum communication. QSD is usually scenarized as follows. Two communicating parties Alice and Bob agree on a set of quantum states $\{|\psi_1\rangle, |\psi_2\rangle,\cdots, |\psi_n\rangle\}$, which correspond to alphabet $\{x_1, x_2, \cdots, x_n \}$ with prior probabilities of each state publicly known. Then Alice encodes the message in the states which are subsequently sent to Bob. Bob decodes the message by discriminating the received states \cite{Bae_2015}. In Hermitian quantum mechanics, for a set of orthogonal quantum states, Bob can discriminate the states with a single copy by using a projective measurement. However, for nonorthogonal quantum states, Bob cannot discriminate them with a single copy because of the collapse of quantum states. 

Much attention has been paid to the discrimination of nonorthogonal quantum states. The minimum error discrimination (MED) \cite{helstrom1969quantum, HOLEVO1973337, solis2017experimental} and the unambiguous state discrimination (USD) \cite{clarke2001experimental, mohseni2004optical, agnew2014discriminating} are the two most investigated strategies in Hermitian systems. In MED, nonorthogonal quantum states are projected onto an orthogonal basis and the result is determined by the best guess according to the measurement result. The strategy aims at minimizing the guessing error. In USD, one expands the space of nonorthogonal quantum states to a higher one by utilizing an auxiliary system, then projects the composite states onto an orthogonal basis in the expanded space. The result is conclusive with some probability. So far solutions of optimal MED and USD strategies in Hermitian systems are confined to a specific set of nonorthogonal quantum states \cite{PhysRevA.64.012303}. A universal optimal solution for the discrimination of \textit{arbitrary} nonorthogonal quantum states is still demanding. 

Problems, which are difficult to resolve in Hermitian systems, may find solutions in non-Hermitian systems. $\mathcal{PT}$-symmetric non-Hermitian systems have been a hot topic since they were proposed \cite{PhysRevLett.80.5243,el2018non,christodoulides2018parity,ozdemir2019parity}. In a $\mathcal{PT}$-symmetric system, the condition of the Hermiticity of the Hamiltonian is replaced by the condition that the Hamiltonian commutes with the joint $\mathcal{PT}$ operator, i.e., $[\mathcal{H}, \mathcal{PT}]=0$. Here, $\mathcal{P}$ is the parity reflection operator while $\mathcal{T}$ is the time-reversal operator. The eigenvalues of the Hamiltonian remain real in the $\mathcal{PT}$ symmetry-unbroken regime despite of the non-Hermiticity. $\mathcal{PT}$-symmetric systems have been realized in both classical and quantum systems \cite{PhysRevLett.106.093902,regensburger2012parity,2014PhysRevLett.113.053604,peng2014parity,gao2015observation,PhysRevLett.117.110802,PhysRevX.8.031079,break2020}. Meanwhile, critical phenomena have been observed \cite{PhysRevA.91.052113}, such as increase of entanglement \cite{PhysRevA.90.054301}, information retrieval \cite{PhysRevLett.119.190401, PhysRevLett.123.230401, PhysRevLett.124.230402}, coherence backflow \cite{fang2021experimental,PhysRevA.103.L020201}, chiral population transfer \cite{xu2016topological, doppler2016dynamically} and decoherence dynamics \cite{PhysRevA.94.040101}. 

It was shown in \cite{PhysRevLett.98.040403} that given an initial state $|\psi_I\rangle$ and a final state $|\psi_F\rangle$, the time required for the state $|\psi_I\rangle$ evolving into the state $|\psi_F\rangle$ was finite and nonzero in Hermitian systems, which is a quantum analogue to the classical brachistochrone problem \cite{PhysRevLett.96.060503, PhysRevX.11.011035}. However, the evolution time can be vanishing in $\mathcal{PT}$-symmetric systems \cite{PhysRevLett.98.040403}. The quantum brachistochrone was experimentally investigated in an NMR system where the qubit was prescribed to evolve from the initial state $|0\rangle$ to the final state $|1\rangle$, and the phenomenon of the evolution time vanishing was observed \cite{zheng2013observation}. The paper \cite{2008} theoretically extended the quantum brachistochrone to the $\mathcal{PT}$ symmetry-broken regime and showed the same intriguing feature as predicted in \cite{PhysRevLett.98.040403}. 

In \cite{bender2013pt}, a $\mathcal{PT}$-symmetric Hamiltonian was used to achieve QSD for two nonorthogonal states. It was theoretically shown that nonorthogonal states could evolve into orthogonal states under a $\mathcal{PT}$-symmetric Hamiltonian. The required evolution time approaches zero at the exceptional point \cite{ozdemir2019parity}, which is subjected to the energy constraint that the energy difference between the largest and the smallest eigenvalues of the Hamiltonian is held fixed. In some sense, the $\mathcal{PT}$-symmetric QSD is equivalent to the USD strategy since both give a conclusive result \cite{bender2013pt}. In \cite{balytskyi2021mathcalptsymmetric}, the $\mathcal{PT}$-symmetric QSD was extended to discriminating arbitrary three nonorthogonal states. The procedure is akin to discriminating two nonorthogonal states. The state, with the highest prior probability, evolved to the one that is orthogonal to the other two states and then was unambiguously discriminated from the other two states. Then the other two states were distinguished in the same way as in \cite{bender2013pt}. 

Although the $\mathcal{PT}$-symmetric QSD was previously studied in theory, \textit{an experimental investigation is still absent.} In this paper, we experimentally demonstrate \textcolor{black}{the QSD} in a $\mathcal{PT}$-symmetric system, which is realized by using a \textcolor{black}{lossy} linear optical setup. In our experiment, we allow quantum states to evolve under a $\mathcal{PT}$-symmetric Hamiltonian, and the time-evolution operator is constructed with optical elements.  The contribution of this work is trifold. First, we demonstrate \textcolor{black}{the QSD} in a $\mathcal{PT}$-symmetric system. We observe that the time required for unambiguously discriminating the nonorthogonal states decreases as the matrix elements of the $\mathcal{PT}$-symmetric Hamiltonian become large. The time can even be vanishingly small when the matrix elements of the Hamiltonian grow and diverge. Second, we find that depending on the overlap of the initial nonorthogonal states, the time-evolved states will not become orthogonal in some regions, which means that they cannot be unambiguously discriminated. At the critical value, the $\mathcal{PT}$-symmetric QSD is equivalent to the optimal USD strategy in Hermitian systems. Third, \textcolor{black}{we observe that the cost of the  discrimination is a loss of photons, whereas more information can be obtained through the measurement as the system tends to the exceptional point.} To the best of our knowledge, our work is the first to observe \textcolor{black}{the QSD} in a $\mathcal{PT}$-symmetric system and also the first to explore the relation between the $\mathcal{PT}$-symmetric QSD and the QSD in Hermitian systems. 

\textcolor{black}{This paper is organized as follows. In section II, we provide the theory and experiment of $\mathcal{PT}$-symmetric QSD for two-state discrimination. The case of three-state discrimination is presented in section III. In section IV, we summarize and discuss this work.}

\begin{figure*}[tbp!]
\centering\includegraphics[width=0.8\textwidth]{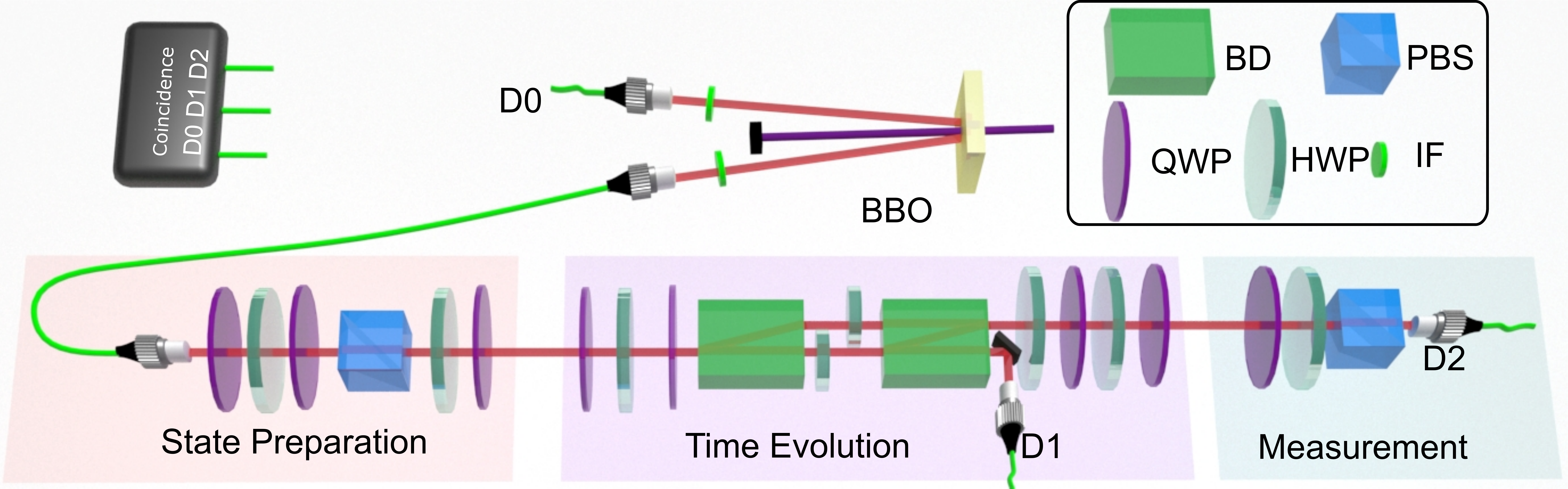}
\caption{Sketch of the experiment. 
The idler photon is detected by D0 for coincidence counting. The signal photon enters the setup, which consists of three parts: state preparation, time evolution, and state measurement. The photon loss is collected by D1. BBO: $\beta$-barium-borate, PBS: polarization beam splitter, BD: beam displacer, HWP: half-wave plate, QWP: quarter-wave plate, IF: interference filter, D0, D1 and D2 are single photon detectors.}
\label{fig1}
\end{figure*}
\section{Two-state discrimination}
\subsection{Theory}
Without loss of generality, we consider two nonorthogonal quantum states $|\psi_1\rangle$ and $|\psi_2\rangle$ in a two-dimensional Hilbert space with overlap $\langle\psi_1|\psi_2\rangle=\cos\epsilon$, parameterized on the Bloch sphere
\begin{eqnarray}
|\psi_1\rangle=\begin{pmatrix}
\cos\frac{\pi-2\epsilon}{4} \\
-i\sin\frac{\pi-2\epsilon}{4} \end{pmatrix},\quad |\psi_2\rangle=\begin{pmatrix}
\cos\frac{\pi+2\epsilon}{4} \\
-i\sin\frac{\pi+2\epsilon}{4} \end{pmatrix},
\label{state}
\end{eqnarray}
where $\epsilon\in(0,\pi/2)$. Note that any two pure states with an overlap of $\cos\epsilon$ can be transformed into the two nonorthogonal states $|\psi_1\rangle$ and $|\psi_2\rangle$ through unitary operations. In Hermitian systems, one can apply the optimal MED and USD strategies to discriminate the states $|\psi_1\rangle$ and $|\psi_2\rangle$ readily. In non-Hermitian systems, two approaches are feasible to discriminate $|\psi_1\rangle$ and $|\psi_2\rangle$ unambiguously.
One approach finds a $\mathcal{PT}$-symmetric Hamiltonian which defines a new Hilbert space, whose inner product interprets the states $|\psi_1\rangle$ and $|\psi_2\rangle$ as being orthogonal.
Another approach is to find a $\mathcal{PT}$-symmetric Hamiltonian under which the states $|\psi_1\rangle$ and $|\psi_2\rangle$ evolve into orthogonal states \cite{bender2013pt,balytskyi2021mathcalptsymmetric}. In this work, we follow the latter approach. 

A general $\mathcal{PT}$-symmetric Hamiltonian for a two-level system has the following form
\begin{equation}
\mathcal{H}_{\mathcal{PT}}=\left(\begin{array}{ll}
re^{i\theta} & s\\
s & re^{-i\theta}
\end{array}\right)=r\cos\theta\bm{1}+\bm{\sigma}\cdot(s,0,ir\sin\theta),\label{Eq:Hpt}
\end{equation}
where the parameters $r, s$ and $\theta$ are real, $\bm{1}$ is the identity matrix, and $\bm{\sigma}$ are the Pauli matrices. The eigenvalues of $\mathcal{H}_{\mathcal{PT}}$ are given by 
\begin{eqnarray}
E_{\pm}=r\cos\theta\pm\sqrt{s^2-r^2\sin^2\theta},
\end{eqnarray}
which are real numbers provided $\sin\alpha=(r\sin\theta)/s<1$ (the $\mathcal{PT}$ symmetry-unbroken regime). The energy constraint indicates that the difference between the eigenvalues 
\begin{eqnarray}
2\omega&=&E_+-E_-\\
&=&2\sqrt{s^2-r^2\sin^2\theta}
\end{eqnarray}
is a constant. The time-evolution operator governed by $\mathcal{H}_{\mathcal{PT}}$ is
\begin{eqnarray}
\bm{U}_{\mathcal{PT}}(t)
&=&\frac{e^{-irt\cos\theta}}{\cos\alpha}\left[\begin{array}{ll}
\cos(\omega t-\alpha) & -i\sin(\omega t)\\
-i\sin(\omega t) & \cos(\omega t+\alpha)
\end{array}\right],\label{Eq:Upt}
\end{eqnarray}
where we set $\hbar=1$.
For the two initial nonorthogonal states $|\psi_1\rangle$ and $|\psi_2\rangle$, the inner product of their time-evolved states under $\bm{U}_{\mathcal{PT}}$ is given by
\begin{eqnarray}
&&\langle\psi_1|\bm{U}_{\mathcal{PT}}^{\dagger}\bm{U}_{\mathcal{PT}}|\psi_2\rangle \nonumber \\
&=&\frac{2\sin^2(\omega t)(\sin^2\alpha\cos\epsilon-\sin\alpha)+\cos\epsilon\cos^2(\alpha)}{\cos^2\alpha},
\end{eqnarray}
which vanishes when 
\begin{eqnarray}
\sin^2(\omega t)=\frac{\cos^2\alpha\cos\epsilon}{2\sin\alpha-2\sin^2\alpha\cos\epsilon}.
\label{eqt}
\end{eqnarray}
The positivity of the right-hand side of Eq.~(\ref{eqt}) is guaranteed by the conditions (i) $\sin\alpha<1$ and (ii) $0<\epsilon<\pi/2$. A solution of Eq.~(\ref{eqt}) gives
\begin{eqnarray}
t=t_0, \pi-t_0,
\label{eqt1}
\end{eqnarray}
with 
\begin{equation}
t_0={\rm arcsin}\left(\sqrt{\frac{\cos^2\alpha\cos\epsilon}{2\sin\alpha-2\sin^2\alpha\cos\epsilon}}\right).
\end{equation}
 A nontrivial solution of $t$ requires $\cos\epsilon\leq 2\sin\alpha/(1+\sin^2\alpha)$. Equation (\ref{eqt1}) indicates that the time-evolved states $\bm{U}_{\mathcal{PT}}|\psi_1\rangle$ and $\bm{U}_{\mathcal{PT}}|\psi_2\rangle$ become orthogonal twice in one period. At the critical value where
  \begin{equation}
\cos\epsilon=2\sin\alpha/(1+\sin^2\alpha),
\label{cv}
  \end{equation}
we have $t_0=\pi/2$. Thus, the two times $t_0$ and $\pi-t_0$ coincide. Note that, in the limit when $\cos\alpha\rightarrow 0$ (the exceptional point), we have $t_0\rightarrow 0$. In this case, we have $s \rightarrow \infty$ and $r\rightarrow \infty$ because $\omega=s\cos\alpha$ is fixed. For $s \rightarrow \infty$ and $r\rightarrow \infty$, one can see from Eq.~(\ref{Eq:Hpt}) that the matrix elements of the Hamiltonian $\mathcal{H}_{\mathcal{PT}}$ tend to $\infty$. Thus we have a vanishing $t_0$ when the matrix elements of $\mathcal{H}_{\mathcal{PT}}$ become infinite.


\subsection{Experimental results}
The subject of simulating quantum state evolutions in a two-dimensional Hilbert space is relevant to a qubit. In the experiment, we utilize a polarized single-photon as the qubit, with $|H\rangle=(1, 0)^{T}$ and $|V\rangle=(0, 1)^{T}$. The time evolution of the qubit is simulated as the photon's polarization state undergoing some optical elements in an optical setup. Single photons have been widely used in simulating a non-Hermitian system to study various critical phenomena \cite{tang2016experimental,Li:19,PhysRevLett.123.230401,PhysRevLett.124.230402,fang2021experimental}. In this work, the single-photon source is generated through a spontaneous parametric down-conversion process by pumping a type-I phase-matched nonlinear $\beta$-barium-borate crystal with a 404 nm pump laser. \textcolor{black}{The power of the pump laser is 130~mW.} The single photon is filtered by an interference filter with bandwidth 10 nm, which yields an average count of 30,000 per second. The idler photon is detected by a single-photon detector for coincidence counting. The signal photon enters the setup which consists of three parts: state preparation, time evolution, and state measurement, as shown in Fig.~\ref{fig1}.

In the state preparation, a combination of wave plates and a polarization beam splitter prepares the initial polarization of the photon to be horizontal with maximum probability. Subsequently, a half-wave plate oriented at $(\pi\pm 2\epsilon)/8$, together with a quarter-wave plate oriented at $\pi/2$, prepares the initial states $|\psi_{1}\rangle$ and $|\psi_{2}\rangle$.

In the time evolution part, the time-evolution operator is decomposed into product of unitary operators and a loss-dependent operator by the singular value decomposition \cite{PhysRevX.8.021017}. The unitary operators are realized by the combination of wave plates \cite{PhysRevA.85.022323}. The loss-dependent operator represents the dissipation of quantum states into the environment, which is realized by a polarization interferometer (consisting of two beam displacers) and a half-wave plate inside.

\begin{figure*}[tbp!]
\centering\includegraphics[width=1\textwidth]{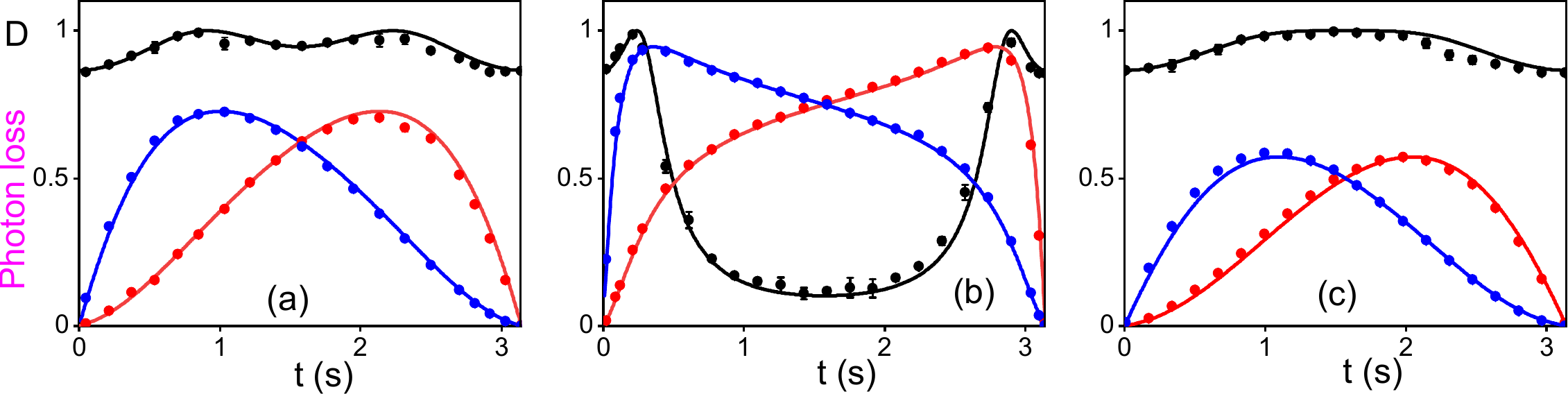}
\caption{Time evolution of the states when $\epsilon=\pi/3$ with (a) $s=1.1$, (b) $s=3$ and (c) $s=1.038$. The black lines are the dynamics of the trace distance $D$ between the time-evolved states $\bm{U}_{\mathcal{PT}}|\psi_1\rangle$ and $\bm{U}_{\mathcal{PT}}|\psi_2\rangle$. The red (blue) lines indicate the dissipations of $|\psi_1\rangle$ ($|\psi_2\rangle$) into the environment. Dots with error bars are the experimental data while lines are the theoretical simulations.}
\label{fig2}
\end{figure*}
\begin{figure*}[tbp!]
\centering\includegraphics[width=1\textwidth]{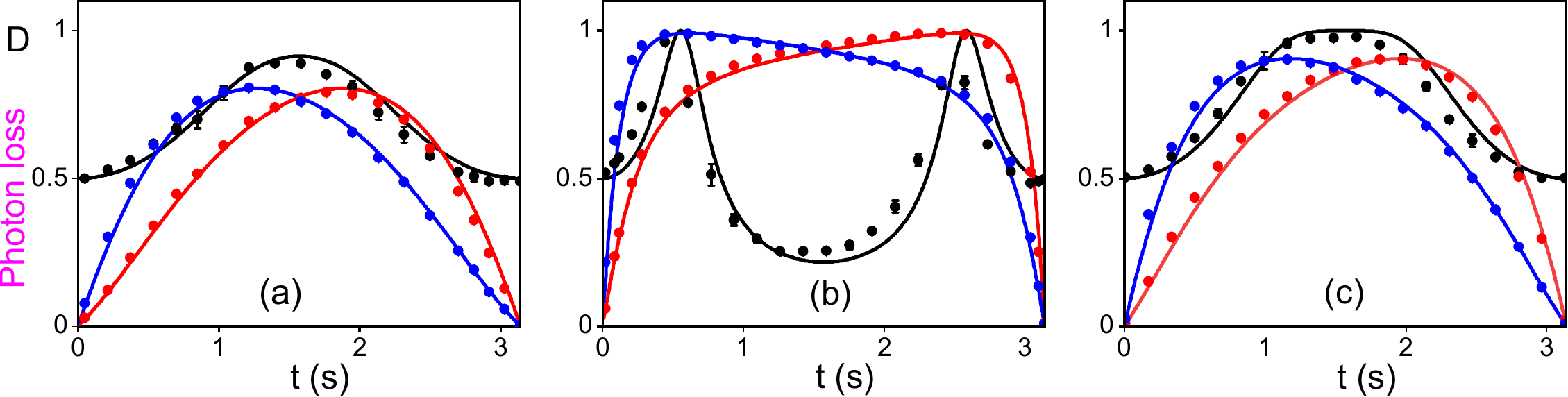}
\caption{Time evolution of the states when $\epsilon=\pi/6$ with (a) $s=1.1$, (b) $s=3$ and (c) $s=1.225$. The black lines are the dynamics of the trace distance $D$ between the time-evolved states $\bm{U}_{\mathcal{PT}}|\psi_1\rangle$ and $\bm{U}_{\mathcal{PT}}|\psi_2\rangle$. The red (blue) lines indicate the dissipations of $|\psi_1\rangle$ ($|\psi_2\rangle$) into the environment. Dots with error bars are the experimental data while lines are the theoretical simulations.}
\label{fig3}
\end{figure*}

The interference at the beam displacers has a visibility higher than $99\%$. We access the time-evolved states ($\bm{U}_{\mathcal{PT}}|\psi_1\rangle$ and $\bm{U}_{\mathcal{PT}}|\psi_2\rangle$) by enforcing the time-evolution operator $\bm{U}_{\mathcal{PT}}$ on the initial states $|\psi_1\rangle$ and $|\psi_2\rangle$ at the specific time. \textcolor{black}{Note that the Hamiltonian in Eq.~(\ref{Eq:Hpt}) is usually employed to describe a system with balanced gain and loss, while our optical setup to simulate the nonunitary time-evolution operator is lossy. Using our setup to simulate a system with gain and loss is achieved by excluding a scale factor of the corresponding time-evolution operator to eliminate the gain of the system (see Appendix A).} In the measurement part, we perform quantum state tomography to reconstruct the density matrices of the time-evolved states.

To observe \textcolor{black}{the QSD}, we fix $\omega=1$ in our experiment, which represents the constant energy difference of the eigenvalues of the non-Hermitian Hamiltonian $\mathcal{H}_{\mathcal{PT}}$. \textcolor{black}{We also set $\theta=\pi/2$, which turns the real parts of the diagonal entries of the Hamiltonian to zero.} We experimentally investigate the time evolution of the states $|\psi_1\rangle$ and $|\psi_2\rangle$ in one period, i.e., the period from $t=0$ to $t=\pi$. To quantify the distinguishability between the time-evolved states, we adopt the trace distance defined by
\begin{eqnarray}
D(\rho_1,\rho_2)=\frac{1}{2}{\rm tr}|\rho_1-\rho_2|,
\end{eqnarray}
with $|A|=\sqrt{A^{\dagger}A}$. Here, $\rho_1$ and $\rho_2$ are the density matrices of the time-evolved states $\bm{U}_{\mathcal{PT}}|\psi_1\rangle$ and $\bm{U}_{\mathcal{PT}}|\psi_2\rangle$, respectively. 
In our work, the dissipation is defined by the loss in photon number $n_1/(n_1+n_2)$, where $n_i$ is the count of $D_i$.

We experimentally investigate two sets of initial states $|\psi_1\rangle$ and $|\psi_2\rangle$, with $\epsilon=\pi/3$ (Fig.~\ref{fig2}) and $\epsilon=\pi/6$ (Fig.~\ref{fig3}), respectively.
For $\epsilon=\pi/3$, a requirement for a nontrivial solution of $t_0$ is $s\geq1.038$.

In Fig.~\ref{fig2}, we set (a) $s=1.1$, (b) $s=3$, and (c) $s=1.038$. The \textcolor{black}{black lines are} the distinguishability between the time-evolved states. The \textcolor{black}{red (blue) lines are} the dissipation of the state $|\psi_1\rangle$ ($|\psi_2\rangle$) into the environment. For $s=1.1$ and $s=3$, there are two points where the time-evolved states $\bm{U}_{\mathcal{PT}}|\psi_1\rangle$ and $\bm{U}_{\mathcal{PT}}|\psi_2\rangle$ become orthogonal, i.e., $D=1$, corresponding to the time $t_0$ and the time $\pi-t_0$. At the critical value $s=1.038$, a solution for $t$ is $t=\pi/2$, where the dissipations of $|\psi_1\rangle$ and $|\psi_2\rangle$ are equal.

From Fig.~\ref{fig2}, one can see that as $s$ increases, the time $t_0$ decreases and the dissipation of  $|\psi_1\rangle$ ($|\psi_2\rangle$) increases. Also, the dissipations are complementary. At time $t_0$, the state $|\psi_1\rangle$ suffers from a larger loss; while at time $\pi-t_0$, the state $|\psi_2\rangle$ suffers from a larger loss. Since the dissipation is straightforwardly related to the photon counts, the variance of the dissipation is smaller than that of the reconstructed density matrix, thus the error bar of the dissipation is smaller than that of the distinguishability.

In Fig.~\ref{fig3}, we set $\epsilon=\pi/6$, for which a requirement for a nontrivial solution of $t_0$ is $s\geq1.225$. We set (a) $s=1.1$, (b) $s=3$, and (c) $s=1.225$. In Fig.~\ref{fig3}(a), the nonorthogonal states $|\psi_1\rangle$ and $|\psi_2\rangle$ never evolve into orthogonal states. In Fig.~\ref{fig3}(b) and Fig.~\ref{fig3}(c), both distinguishability and dissipations exhibit similar patterns as those in Fig.~\ref{fig2}(b) and Fig.~\ref{fig2}(c). For $\epsilon=\pi/6$ and the same $s$, it takes more time for the two nonorthogonal states $|\psi_1\rangle$ and $|\psi_2\rangle$ to evolve into two orthogonal states. In this case, since the overlap of the initial states $|\psi_1\rangle$ and $|\psi_2\rangle$ is larger, their dissipations into the environment are also larger.

\begin{figure}[tbp!]
\centering\includegraphics[width=0.5\textwidth]{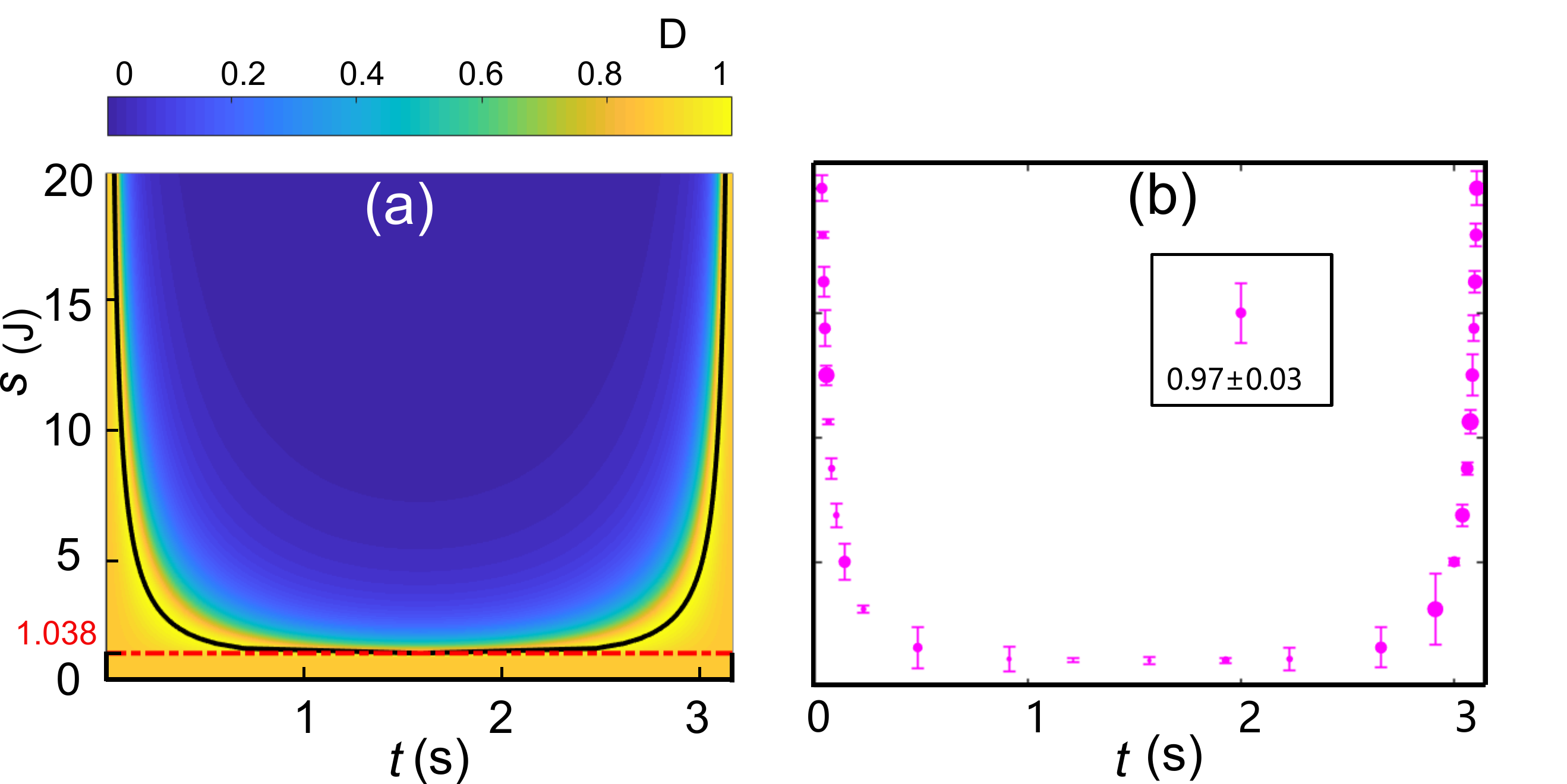}
\caption{(a) Theoretical simulation of the dynamics of $D$ for different values of $s$. The black line indicates the points where the two nonorthogonal states $|\psi_1\rangle$ and $|\psi_2\rangle$ evolve into the orthogonal states. (b) Experimentally measured $D$ at the points where $|\psi_1\rangle$ and $|\psi_2\rangle$ evolve into orthogonal states. The sizes of the dots and the length of the error bars are proportional to the deviation from the theoretical value (unity) and the standard deviation of $D$, respectively. The inset shows a typical value of $0.97\pm 0.03$.}
\label{fig4}
\end{figure}
\begin{figure}[tbp!]
\centering\includegraphics[width=0.48\textwidth]{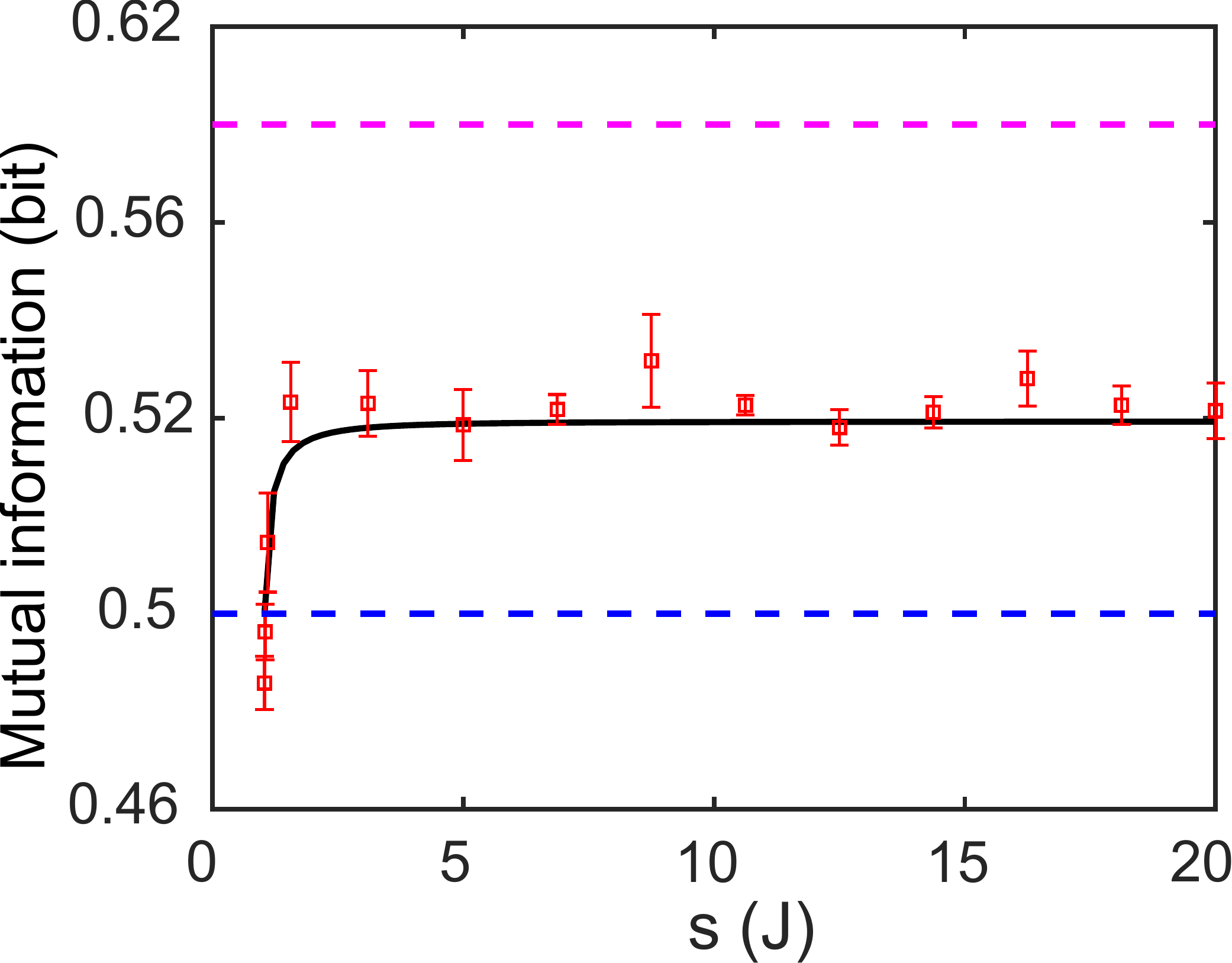}
\caption{Mutual information obtained for the case of  $\epsilon=\pi/3$. The horizontal blue (magenta) dashed line is the mutual information obtained through the optimal USD (MED) strategy. The black curve is the mutual information obtained in the $\mathcal{PT}$-symmetric QSD. Dots with error bars are experimental values. Note that the black curve intersects the horizontal dashed blue line at $s\approx 1.038$.}
\label{fig5}
\end{figure}

To further study the phenomenon of \textcolor{black}{QSD}, we investigate the dynamics of distinguishability under different values of $s$ for $\epsilon=\pi/3$. Figure \ref{fig4}(a) shows our theoretical simulation.  For each $s$, there are two times, $t_0$ and $\pi-t_0$, where the time-evolved states $\bm{U}_{\mathcal{PT}}|\psi_1\rangle$ and $\bm{U}_{\mathcal{PT}}|\psi_2\rangle$ become orthogonal. The black U-shaped curve indicates the positions where $D=1$ in one period from $t=0$ to $t=\pi$. At the critical value $s=1.038$, the two times $t_0$ and $\pi-t_0$ coincide. The $t_0$ tends to zero as $s$ tends to infinity. Figure \ref{fig4}(b) shows the U-shaped experimental distinguishability at times $t_0$ and $\pi-t_0$ for each $s$. The sizes of the dots and the length of the error bars are proportional to the deviation from the theoretical value (unity) and the standard deviation of $D$, respectively. For reference, the inset shows a typical value of $0.97\pm 0.03$.

Finally, we compare the $\mathcal{PT}$-symmetric QSD with the optimal MED and USD strategies in Hermitian systems. Let the prior probabilities of $|\psi_1\rangle$ and $|\psi_2\rangle$ be equal. We measure the mutual information obtained through measurement (see Appendix B). Figure \ref{fig5} shows the mutual information under different $s$ for  $\epsilon=\pi/3$. The black line is the theoretical prediction of the mutual information obtained in the $\mathcal{PT}$-symmetric QSD. The blue dashed line (0.5) and the magenta dashed line (0.6) are the mutual informations obtained by using the optimal USD and MED strategies, respectively. As $s$ increases, the amount of information, obtained by the $\mathcal{PT}$-symmetric QSD, increases and tends to be constant. Also, the black curve and the horizontal blue dashed line intersect at the point $s=1.038$, which means that at that point the $\mathcal{PT}$-symmetric QSD and the optimal USD strategy induce the same amount of information. Therefore, at that critical value, using a $\mathcal{PT}$-symmetric Hamiltonian to perform the QSD is equivalent to applying the optimal USD strategy.

\section{Three-state discrimination}
\subsection{Theory}
{\color{black}
The $\mathcal{PT}$-symmetric Hamiltonian can also be applied to discriminate three nonorthogonal arbitrary states \cite{balytskyi2021mathcalptsymmetric}. Without loss of generality, we consider three nonorthogonal arbitrary quantum states
\begin{equation}
|\psi_j\rangle=\begin{pmatrix}
\cos\frac{\beta_j}{2} \\
e^{i\gamma_j}\sin\frac{\beta_j}{2} \end{pmatrix}, \quad j=1,2,3,
\label{31}
\end{equation}
where $\beta_j$ are the parallels and $\gamma_j$ are the meridians of the positions for the state vector of $|\psi_j\rangle$ on the Bloch sphere. The procedure is to first discriminate one state from the other two in the first measurement, and then discriminate the other two states in the second measurement. Therefore, one needs at most two measurements for the three-state discrimination.

\begin{figure*}[tbp!]
\centering\includegraphics[width=1\textwidth]{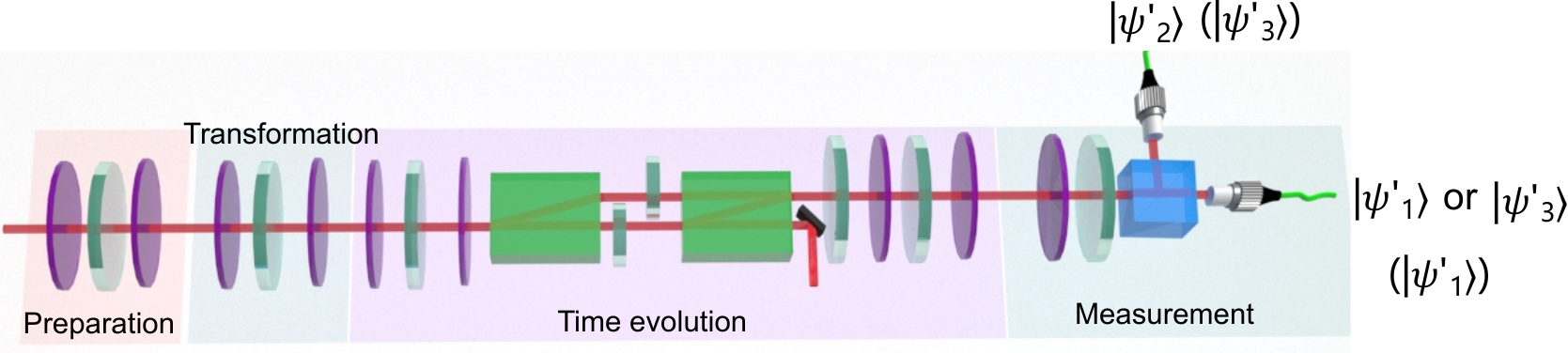}
\caption{Experimental setup for the three-state discrimination. The initial state preparation and the state transformation are realized by combinations of wave plates. The time-evolved states are projected onto the $|H\rangle$ and $|V\rangle$ basis. A detection of the state $|V\rangle$ in the first measurement indicates that the initial state is $|\psi_2\rangle$, otherwise it is $|\psi_1\rangle$ or $|\psi_3\rangle$. In the second measurement, a detection of the state $|V\rangle$ indicates that the initial state is $|\psi_3\rangle$, otherwise it is $|\psi_1\rangle$. The time-evolved states are defined as $|\psi'_1\rangle$, $|\psi'_2\rangle$ and $|\psi'_3\rangle$, which correspond to the three initial states $|\psi_1\rangle$, $|\psi_2\rangle$ and $|\psi_3\rangle$, respectively.}
\label{fig6}
\end{figure*}

To begin with, note that an arbitrary set of three nonorthogonal states can be transformed into the following forms through unitary operations (see Appendix C)
\begin{eqnarray}
\label{3state1}
|\psi_1\rangle&=&\begin{pmatrix}
\cos\frac{\pi-2\epsilon_{12}}{4} \\
-i\sin\frac{\pi-2\epsilon_{12}}{4} \end{pmatrix},\\
\label{3state2} |\psi_2\rangle&=&\begin{pmatrix}
\cos\frac{\pi+2\epsilon_{12}}{4} \\
-i\sin\frac{\pi+2\epsilon_{12}}{4} \end{pmatrix},\\
|\psi_3\rangle&=&\begin{pmatrix}
\cos\frac{\mu}{2} \\
e^{i\nu}\sin\frac{\mu}{2} \end{pmatrix},
\label{3state3}
\end{eqnarray}
with additional overall phases dropped off. Such forms are convenient to transform the states $|\psi_1\rangle$ and $|\psi_2\rangle$ into orthogonal states through a $\mathcal{PT}$-symmetric time evolution according to the previous section. Here, $\cos\epsilon_{12}=|\langle\psi_2|\psi_1\rangle|$ is the overlap between the states $|\psi_1\rangle$ and $|\psi_2\rangle$, $\nu$ and $\mu$ are the meridian and parallel angles of the state $|\psi_3\rangle$, respectively. 

Therefore, if one chooses a $\mathcal{PT}$-symmetric Hamiltonian given by Eq.~(\ref{Eq:Hpt}) with the corresponding evolution time $t$ satisfying Eq.~(\ref{eqt}), the time-evolved states of $|\psi_1\rangle$ and $|\psi_2\rangle$ become orthogonal, while the overlap between the time-evolved states of $|\psi_2\rangle$ and $|\psi_3\rangle$ also decreases as $\alpha$ tends to $\frac{\pi}{2}$ (see Appendix C). One could discriminate $|\psi_2\rangle$ from $|\psi_1\rangle$ and $|\psi_3\rangle$ when $\alpha\rightarrow \frac{\pi}{2}$ in the first measurement. 

In the second measurement, one can transform the states $|\psi_1\rangle$ and $|\psi_3\rangle$ into the similar forms given by Eqs.~(\ref{3state1}) and (\ref{3state2}) 
\begin{eqnarray}
\label{3statea}
|\psi_1\rangle&=&\begin{pmatrix}
\cos\frac{\pi-2\epsilon_{13}}{4} \\
-i\sin\frac{\pi-2\epsilon_{13}}{4} \end{pmatrix},\\
 |\psi_3\rangle&=&\begin{pmatrix}
\cos\frac{\pi+2\epsilon_{13}}{4} \\
-i\sin\frac{\pi+2\epsilon_{13}}{4} \end{pmatrix},
\label{3stateb}
\end{eqnarray}
where $\cos\epsilon_{13}=|\langle\psi_3|\psi_1\rangle|$ is the overlap between $|\psi_1\rangle$ and $|\psi_3\rangle$. Note that since the state $|\psi_2\rangle$ is already excluded in the first measurement, it is ignored in the second measurement. Then the states $|\psi_1\rangle$ and $|\psi_3\rangle$ can be unambiguously discriminated through a $\mathcal{PT}$-symmetric time evolution.

\subsection{Experimental results}
In the experiments, we set $\beta_j=\beta$, $\gamma_j=\frac{2\pi}{3}(j-1)$, i.e., the three states given by Eq.~(\ref{31}) are uniformly located on the circle of the Bloch sphere with parallel angle $\beta$. Figure \ref{fig6} shows our experimental setup. The initial states are first prepared through a combination of wave plates and then transformed into the forms given by Eqs.~(\ref{3state1}-\ref{3state3}) in the first measurement or Eqs.~(\ref{3statea}-\ref{3stateb})  in the second measurement. After that, a time-evolution operator is imposed on the states. In the measurement part, the time-evolved states are projected onto the states $|H\rangle$ and $|V\rangle$ and detected by single-photon detectors.

In this setup, if one detects the state $|V\rangle$ in the first measurement, one asserts that the input state is $|\psi_2\rangle$, otherwise, it is $|\psi_1\rangle$ or $|\psi_3\rangle$. Then in the second measurement, the states $|\psi_1\rangle$ and $|\psi_3\rangle$ are discriminated unambiguously. Note that in the first measurement, since the time-evolved states of $|\psi_2\rangle$ and $|\psi_3\rangle$ are not strictly orthogonal, there is a probability that one gets a wrong result (i.e., $|\psi_2\rangle$) if the input state is $|\psi_3\rangle$. Provided the initial state is $|\psi_j\rangle$, the probability $P_j$ of the correct result is given by
\begin{eqnarray}
P_1&=&\frac{N_{1H}}{N_{1H}+N_{1V}}\times\frac{N_{2H}}{N_{2V}+N_{2H}},\\
P_2&=&\frac{N_{1V}}{N_{1H}+N_{1V}},\\
P_3&=&\frac{N_{1H}}{N_{1H}+N_{1V}}\times\frac{N_{2V}}{N_{2V}+N_{2H}},
\end{eqnarray}
where $N_{jH} (N_{jV})$ is the photon count for the state $|H\rangle (|V\rangle)$ in the $j$th measurement.

\begin{figure}[tbp!]
\centering\includegraphics[width=0.48\textwidth]{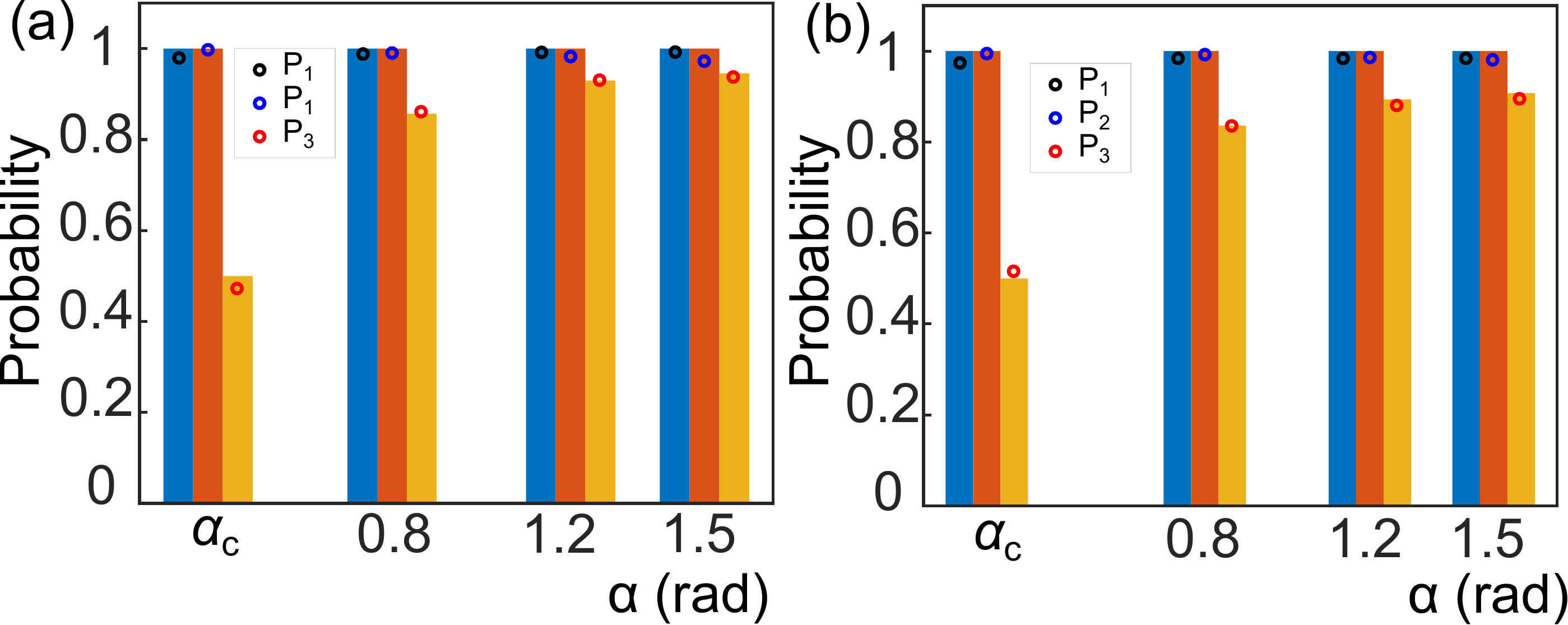}
\caption{Probabilities of correctly finding the states for (a) $\beta=\pi/3$ and (b) $\beta=\pi/2$. The black, blue, and red circles are experimental results when the input states are $|\psi_1\rangle$, $|\psi_2\rangle$, and $|\psi_3\rangle$, respectively. The bars are theoretical values. (a) $\alpha_c=0.39$ and (b) $\alpha_c=0.27$ are the critical values of $\alpha$ given by Eq.~(\ref{cv}). The error bars are not shown because they are too small.}
\label{fig3state}
\end{figure}

Figure \ref{fig3state} shows our experimental results. We set $\alpha=\alpha_c, 0.8, 1.2$ and 1.5, where $\alpha_c$ is the critical value of $\alpha$ given by Eq.~(\ref{cv}). We also set $\beta=\pi/3$ [Fig.~\ref{fig3state}(a)] and $\beta=\pi/2$ [Fig.~\ref{fig3state}(b)]. One can see from Fig.\ref{fig3state} that $P_1$ and $P_2$ are theoretically equal to unity. This implies that if the initial state is $|\psi_1\rangle$ or $|\psi_2\rangle$, one can always get a right result. However, when the initial state is $|\psi_3\rangle$, the probability of the correct result is less than 1 and it increases as $\alpha$ tends to $\pi/2$. This is because the time-evolved states of $|\psi_2\rangle$ and $|\psi_3\rangle$ are not orthogonal and there is a probability that it yields a wrong result. But the overlap between the time-evolved states of $|\psi_3\rangle$ and $|\psi_2\rangle$ becomes smaller as $\alpha$ tends to $\pi/2$. The experimental results show that one needs one measurement to find the state $|\psi_2\rangle$, whereas one needs two measurements to find the states $|\psi_1\rangle$ and $|\psi_3\rangle$. Note that due to the nonunitarity of the time evolution, there exists photon loss in the process of the time evolution. Therefore, though in principle one is able to discriminate the three states with at most two measurements, more than two samples may be required for the state discrimination.
}
%
\section{Conclusion and discussion}
$\mathcal{PT}$-symmetric theory has been well developed since it was put forward \cite{PhysRevLett.112.130404,PhysRevA.78.042115,mostafazadeh2002pseudo1,mostafazadeh2002pseudo2}, and whether a $\mathcal{PT}$-symmetric system can outperform a Hermitian system was also argued \cite{PhysRevA.91.052113,PhysRevLett.99.130502}. In this	 work, we observe the phenomenon of \textcolor{black}{quantum state discrimination} by allowing quantum states to evolve under a $\mathcal{PT}$-symmetric Hamiltonian.
A non-Hermitian Hamiltonian generates a unitary time-evolution in a new Hilbert space, provided the inner product is suitably defined. In the new space, two initial nonorthogonal states may be interpreted as orthogonal states, which means the geodesic distance between the states is different in the new space \cite{PhysRevLett.99.130502}. At the exceptional point, the eigenstates of $\mathcal{H}_{\mathcal{PT}}$ coalesce, which is the cause of many critical phenomena in $\mathcal{PT}$-symmetric systems; thus the time required for nonorthogonal states to evolve into orthogonal states can be close to zero.

In summary, we have experimentally demonstrated \textcolor{black}{the quantum state discrimination for two nonorthogonal states and three nonorthogonal states} in a $\mathcal{PT}$-symmetric system, which is implemented by using a linear optical setup. For two-state discrimination, we have observed that as the matrix elements of the Hamiltonian become large, the time required to discriminate two nonorthogonal states decreases. The time can even vanish when the matrix elements of the Hamiltonian approach infinity. We have also observed that the cost of such a \textcolor{black}{state discrimination is the dissipation of quantum states into the environment}. In addition, we have shown that, at a critical value, $\mathcal{PT}$-symmetric quantum state discrimination is equivalent to the optimal USD strategy in Hermitian systems, both inducing the same amount of mutual information and dissipation. \textcolor{black}{For three-state discrimination, we have shown that at most two measurements are required to find the correct states.} \textcolor{black}{Compared to the previous works, our results reveal more features of the $\mathcal{PT}$-symmetric quantum state discrimination. Moreover, we experimentally demonstrate that the $\mathcal{PT}$-symmetric quantum state discrimination is equivalent to the unambiguous discrimination strategy in Hermitian systems.} This work provides physical insight into $\mathcal{PT}$-symmetric quantum state discrimination, and may promote the application of $\mathcal{PT}$-symmetric theory in quantum information processing and quantum communications \cite{ju2019non}.

\section*{Acknowledgements}
This work is partly supported by the National Natural Science Foundation of China (NSFC) (Grants No. 11774076, 11804228, U21A20436), Jiangxi Natural Science Foundation (Grant No. 20192ACBL20051), and the Key-Area Research and Development Program of GuangDong province (2018B03-0326001). F.N. is supported in part by: Nippon Telegraph and Telephone Corporation (NTT) Research, the Japan Science and Technology Agency (JST) [via the Quantum Leap Flagship Program (Q-LEAP), and the Moonshot R\&D Grant Number JPMJMS2061], the Japan Society for the Promotion of Science (JSPS) [via the Grants-in-Aid for Scientific Research (KAKENHI) Grant No. JP20H00134], the Army Research Office (ARO) (Grant No. W911NF-18-1-0358), the Asian Office of Aerospace Research and Development (AOARD) (via Grant No. FA2386-20-1-4069), and the Foundational Questions Institute Fund (FQXi) via Grant No. FQXi-IAF19-06.


\onecolumngrid
\setcounter{equation}{0} 
\setcounter{figure}{0} 
\renewcommand{\thefigure}{S\arabic{figure}}
\renewcommand{\theequation}{S\arabic{equation}}
\section*{Appendix A: Realization of the nonunitary time-evolution operator}
The nonunitary time-evolution operator of a $\mathcal{PT}$-symmetric Hamiltonian has the following form
\begin{eqnarray}
\bm{U}_{\mathcal{PT}}(t)&=&e^{-i\hat{H}_{\mathcal{PT}}t}=\frac{e^{-irt\cos\theta}}{\cos\alpha}\left[\begin{array}{ll}
\cos(\omega t-\alpha) & -i\sin(\omega t)\\
-i\sin(\omega t) & \cos(\omega t+\alpha)
\end{array}\right],
\end{eqnarray}
which can be decomposed (via the singular-value decomposition \cite{PhysRevX.8.021017}) into a product of unitary operators and a diagonalized loss-dependent operator
\begin{eqnarray}
\bm{U}_{\mathcal{PT}}(t)&=&cTMW,\label{Eq:Upt1}
\end{eqnarray}
\textcolor{black}{where $T$ and $W$ are unitary matrices, $M$ is a diagonal matrix, and $c$ is a scale factor which sets the maximum of the diagonal entries of $M$ to be unity. Note that in our experiment the scale factor $c$ is neglected to eliminate the gain of the system and we experimentally realized the joint operation $TMW$. This is reasonable since the effects of $\bm{U}_{\mathcal{PT}}(t)$ and $TMW$ on the quantum states, after renormalizing the time-evolved states, are the same \cite{PhysRevLett.123.230401}.}

Note that an arbitrary unitary matrix can be parametrized as
\begin{eqnarray}
{u}&=&\left(\begin{array}{ll}
a_0-ia_1 & -ia_2-a_3\\
-ia_2+a_3 & a_0+ia_1
\end{array}\right)\label{Eq:Upt2}
\end{eqnarray}
by multiplying a global phase, where $a_j (j=0,1,2,3)$ are real numbers, $a_0\geq 0$, and $\sum_{j=0}^{j=3} a_j^2=1$. Equation (\ref{Eq:Upt2}) can be decomposed into SU(2) gates on the basis of the Euler angle parametrization \cite{PhysRevA.85.022323}:
\begin{eqnarray}
{u}\equiv \exp(-i\frac{1}{2}\xi\sigma_2)\exp(-i\frac{1}{2}\eta\sigma_3)\exp(-i\frac{1}{2}\zeta\sigma_2),\label{Eq:Upt3}
\end{eqnarray}
where
\begin{eqnarray}
\xi&=&\arctan\left(\frac{a_3}{a_0}\right)+\arctan\left(\frac{a_2}{a_1}\right)+\frac{\pi}{2}(1-{\rm sign}a_1), \\
\eta&=&2\arccos\sqrt{a_0^2+a_3^2},\\
\zeta&=& \arctan\left(\frac{a_3}{a_0}\right)-\arctan\left(\frac{a_2}{a_1}\right)-\frac{\pi}{2}(1-{\rm sign}a_1).
\end{eqnarray}
The unitary matrix ${u}$ can thus be realized by a combination of wave plates
\begin{equation}
{u}=Q_{\pi/4+\xi/2}H_{-\pi/4+(\xi+\eta-\zeta)/4}Q_{\pi/4-\zeta/2},
\label{wp}
\end{equation}
where $Q_{\phi}(H_{\phi})$ is the Jones matrix of a quarter-(half-)wave plate with fast-axis orientation $\phi$. Note that the $u$ considered here is an arbitrary unitary matrix. Thus, according to Eqs.~(\ref{Eq:Upt3}-\ref{wp}), the unitary matrices $T$ and $W$ can be implemented with a combination of wave plates, given that the matrix elements of $T$ and $W$ are known.

\begin{figure}[htbp!]
\centering\includegraphics[width=0.5\textwidth]{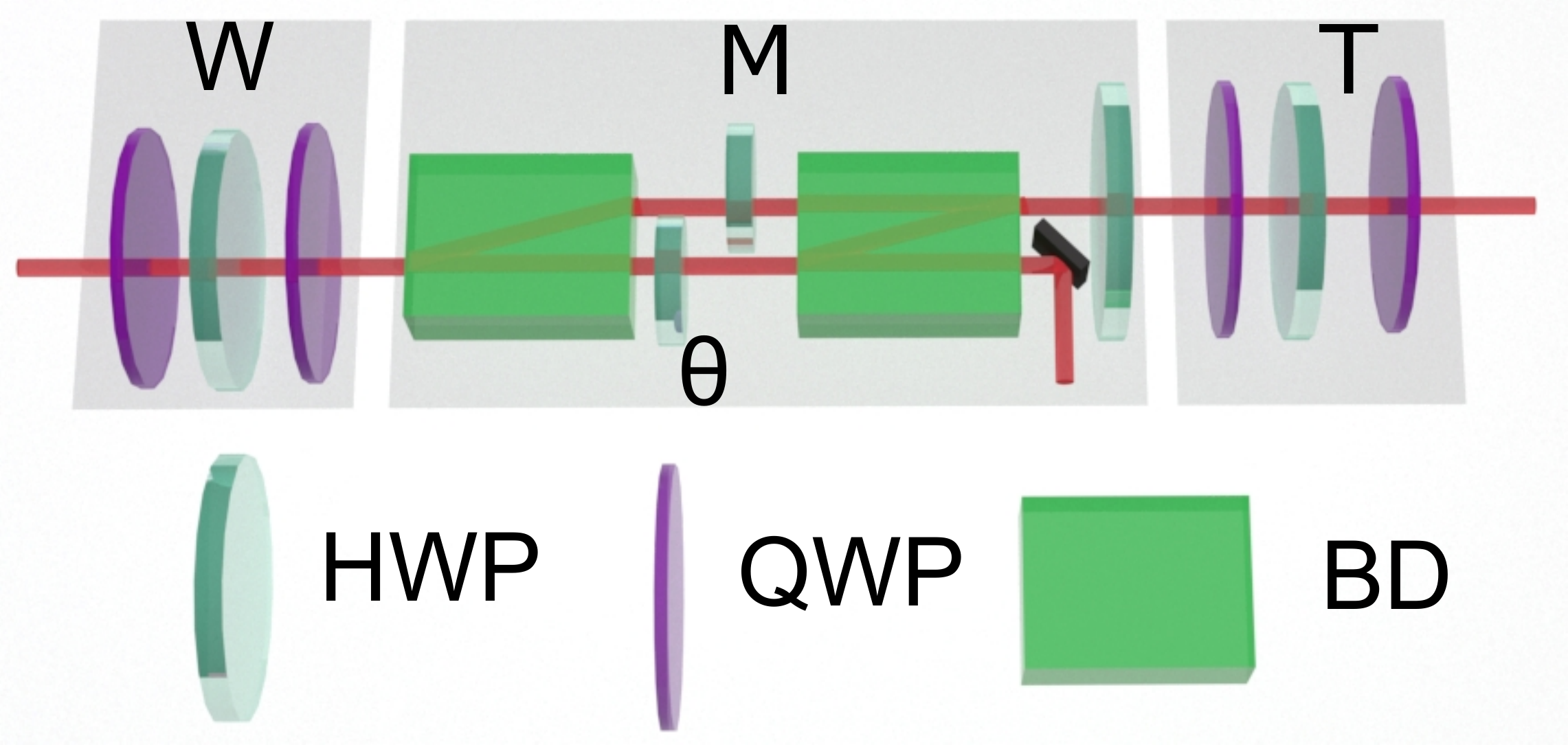}
\caption{Optical simulation of the nonunitary time-evolution operator.}
\label{figs1}
\end{figure}

Moreover, the loss-dependent operator $M$ can be written as \cite{fang2021experimental,PhysRevLett.123.230401}:
\begin{eqnarray}
M=\left[\begin{array}{ll}
1 & 0\\
0 & \sin(2\theta)
\end{array}\right],\label{Eq:Upt4}
\end{eqnarray}
which leaves a H-polarized photon undissipated and attenuates a V-polarized photon by $\sin(2\theta)$. \textcolor{black}{Note that here the maximum of the diagonal entries of $M$ is unity, which is assured by the scale factor $c$ in Eq.~(\ref{Eq:Upt1}).} The operator $M$ is realized by a polarization interferometer (composed of two beam displacers) and a half-wave plate inside (oriented at $\theta$). The beam displacers transmit the V-polarized photon and displace the H-polarized photon. A half-wave plate oriented at $\theta$ is inserted in the V path to induce the dissipation. The optical simulation of $\bm{U}_{\mathcal{PT}}$ is shown in Fig. (\ref{figs1}) where a photon enters $W$, $M$, and $T$ parts in sequence.
\section*{Appendix B: Mutual information}
The mutual information between Alice (A) and Bob (B) is \cite{barnett2009quantum,Bae_2015} 
\begin{eqnarray}
H(A:B)=\sum\limits_{ij}p_i{\rm Tr}(\hat{\rho}_i\hat{\pi}_j)\log\left[ \frac{{\rm Tr}(\hat{\rho}_i\hat{\pi}_j)}{{\rm Tr}(\hat{\rho}\hat{\pi}_j)}\right],
\label{mi}
\end{eqnarray}
where the quantum state $\hat{\rho}_i$ is prepared by Alice with \textit{a priori} probability $p_i$ and Bob performs a positive operator-valued measure $\{\hat{\pi}_j, j=1,2,3\}$, with $\sum_{j}\hat{\pi}_j=\mathbb{I},\quad\hat{\rho}=\sum_{i}p_i\hat{\rho}_i$. Here, we assume the prior probabilities of $|\psi_1\rangle$ and $|\psi_2\rangle$ to be equal, i.e., $p_1=p_2=0.5$. \textcolor{black}{The mutual information quantifies how much information Bob obtains through the measurement with outcomes corresponding to the expectation values of $\{\hat{\pi_j}\}$.}

\textcolor{black}{Note that the measurement in our experiment is not exactly a projective measurement, but rather a time-evolution process accompanied by energy loss. Therefore, the mutual information defined in Eq.~(\ref{mi}) is better viewed in a classical way, i.e., it is the information obtained on the basis of three different outcomes. With this in mind, one could regard the expectation value of $\hat{\pi}_1$ as the photon loss into the environment. That is, ${\rm Tr}(\hat{\rho_1}\hat{\pi_1})$ and ${\rm Tr}(\hat{\rho_2}\hat{\pi_1})$ are the dissipations of the states $|\psi_1\rangle$ and $|\psi_2\rangle$, respectively.}

\textcolor{black}{On the other hand, the expectation values of $\hat{\pi}_2$ and $\hat{\pi}_3$ could be regarded as the unambiguous results while discriminating the time-evolved states $\bm{U}_{\mathcal{PT}}|\psi_1\rangle$ and $\bm{U}_{\mathcal{PT}}|\psi_2\rangle$. To be specific, ${\rm Tr}(\hat{\rho_1}\hat{\pi_2})$ and ${\rm Tr}(\hat{\rho_2}\hat{\pi_3})$ could be regarded as the unambiguous results for correctly deciding that the states are $|\psi_1\rangle$ and $|\psi_2\rangle$, respectively. Since the time-evolved states $\bm{U}_{\mathcal{PT}}|\psi_1\rangle$ and $\bm{U}_{\mathcal{PT}}|\psi_2\rangle$ are orthogonal, a wrong decision does not exist, i.e., ${\rm Tr}(\hat{\rho_1}\hat{\pi_3})={\rm Tr}(\hat{\rho_2}\hat{\pi_2})=0$.}

\onecolumngrid
\section*{Appendix C: Theory of three-state discrimination}
{\color{black}
First, we show that an arbitrary set of three states with the forms given by Eq.~(\ref{31}) can be transformed into the forms in Eqs.~(\ref{3state1}-\ref{3state3}) through unitary operations. 

The first step is to transform the state $|\psi_1\rangle$ into the state $|0\rangle\equiv (1\quad 0)^T$ with the rotation operation
\begin{equation}
R_1=\left(\begin{array}{cc}
\cos\frac{\beta_1}{2} & \sin\frac{\beta_1}{2}e^{-i\gamma_1}\\
-\sin\frac{\beta_1}{2}e^{i\gamma_1} & \cos\frac{\beta_1}{2}
\end{array}\right).
\end{equation}
Then, the meridian angle of the state $|\psi_2\rangle$ is changed to $3\pi/2$ by a second rotation around the $Z$ axis
\begin{equation}
R_2=\left(\begin{array}{cc}
1 & 0\\
0 & -i\exp[-i(\lambda+\gamma_2)]
\end{array}\right),
\end{equation}
where
\begin{eqnarray}
\nonumber\lambda&=&\arctan\left[\frac{\sin\frac{\beta_1}{2}\cos\frac{\beta_2}{2}\sin(\gamma_2-\gamma_1)}{\cos\frac{\beta_1}{2}\sin\frac{\beta_2}{2}-\sin\frac{\beta_1}{2}\cos\frac{\beta_2}{2}\cos(\gamma_2-\gamma_1)} \right]\\
&&-\arctan\left[\frac{\sin\frac{\beta_1}{2}\sin\frac{\beta_2}{2}\sin(\gamma_2-\gamma_1)}{\cos\frac{\beta_1}{2}\cos\frac{\beta_2}{2}-\sin\frac{\beta_1}{2}\sin\frac{\beta_2}{2}\cos(\gamma_2-\gamma_1)} \right].
\end{eqnarray}
Finally, the parallel angles of the states $|\psi_1\rangle$ and $|\psi_2\rangle$ are changed to $\frac{\pi\mp 2\epsilon_{12}}{2}$ by a rotation around $X$ axis
\begin{equation}
R_3	=\left(\begin{array}{cc}
\cos\frac{\pi- 2\epsilon_{12}}{4} & -i\sin\frac{\pi- 2\epsilon_{12}}{4}\\
-i\sin\frac{\pi- 2\epsilon_{12}}{4} & \cos\frac{\pi- 2\epsilon_{12}}{4}
\end{array}\right),
\end{equation}
where
\begin{equation}
\cos\epsilon_{12}=\sqrt{\frac{1+\cos\beta_1\cos\beta_2+\sin\beta_1\sin\beta_2\cos(\gamma_1-\gamma_2)}{2}}
\end{equation}
is the overlap between the states $|\psi_1\rangle$ and $|\psi_2\rangle$. The joint operation $R_3R_2R_1$ transforms the three states into the forms given by Eqs.~(\ref{3state1}-\ref{3state3}) with
\begin{eqnarray}
\cos\frac{\mu}{2}&=&|\kappa_1|,\\
\nu&=&\arctan\left[\frac{Im(\kappa_2)}{Re(\kappa_2)}\right]-\arctan\left[\frac{Im(\kappa_1)}{Re(\kappa_1)}\right],\\
\kappa_1&=&\cos\frac{\beta_1}{2}\cos\frac{\beta_3}{2}\cos\frac{\pi-2\epsilon_{12}}{4}\left[1+\tan\frac{\beta_1}{2}\tan\frac{\pi-2\epsilon_{12}}{4}e^{i(\gamma_1-\gamma_2-\lambda)}\right]\\
&&+\sin\frac{\beta_1}{2}\sin\frac{\beta_3}{2}\cos\frac{\pi-2\epsilon_{12}}{4}e^{i(\gamma_3-\gamma_1)}\left[1-\cot\frac{\beta_1}{2}\tan\frac{\pi-2\epsilon_{12}}{4}e^{i(\gamma_1-\gamma_2-\lambda)}\right],\\
\kappa_2&=&i\cos\frac{\beta_1}{2}\cos\frac{\beta_3}{2}\sin\frac{\pi-2\epsilon_{12}}{4}\left[\tan\frac{\beta_1}{2}\cot\frac{\pi-2\epsilon_{12}}{4}e^{i(\gamma_1-\gamma_2-\lambda)}-1\right]\\
&&-i\sin\frac{\beta_1}{2}\sin\frac{\beta_3}{2}\sin\frac{\pi-2\epsilon_{12}}{4}e^{i(\gamma_3-\gamma_1)}\left[1+\cot\frac{\beta_1}{2}\cot\frac{\pi-2\epsilon_{12}}{4}e^{i(\gamma_1-\gamma_2-\lambda)}\right].
\end{eqnarray}

Thus, the time-evolved states of the three initial states $|\psi_1\rangle$, $|\psi_2\rangle$, and $|\psi_3\rangle$ given by Eqs.~(\ref{31}), become
\begin{eqnarray}
\bm{U}_{\mathcal{PT}}|\psi_1\rangle&=&\frac{e^{-irt\cos\theta}}{\cos\alpha}\begin{pmatrix}
\cos(\omega t-\alpha)\cos\frac{\pi-2\epsilon_{12}}{4}-\sin \omega t\sin\frac{\pi-2\epsilon_{12}}{4} \\
-i\sin \omega t\cos\frac{\pi-2\epsilon_{12}}{4}-i\cos(\omega t+\alpha)\sin\frac{\pi-2\epsilon_{12}}{4} \end{pmatrix},\\
\bm{U}_{\mathcal{PT}}|\psi_2\rangle&=&\frac{e^{-irt\cos\theta}}{\cos\alpha}\begin{pmatrix}
\cos(\omega t-\alpha)\cos\frac{\pi+2\epsilon_{12}}{4}-\sin \omega t\sin\frac{\pi+2\epsilon_{12}}{4} \\
-i\sin \omega t\cos\frac{\pi+2\epsilon_{12}}{4}-i\cos(\omega t+\alpha)\sin\frac{\pi+2\epsilon_{12}}{4} \end{pmatrix},\\
\bm{U}_{\mathcal{PT}}|\psi_3\rangle&=&\frac{e^{-irt\cos\theta}}{\cos\alpha}\begin{pmatrix}
\cos(\omega t-\alpha)\cos\frac{\mu}{2}-i\sin \omega t e^{i\nu}\sin\frac{\mu}{2} \\
-i\sin \omega t\cos\frac{\mu}{2}+\cos(\omega t+\alpha)e^{i\nu}\sin\frac{\mu}{2} \end{pmatrix},
\end{eqnarray}
which could be normalized to the following forms
\begin{eqnarray}
|\psi'_1\rangle&=&\begin{pmatrix}
\cos\frac{\delta}{2} \\
-i\sin\frac{\delta}{2}\end{pmatrix},\\
|\psi'_2\rangle&=&\begin{pmatrix}
\sin\frac{\delta}{2} \\
i\cos\frac{\delta}{2}\end{pmatrix},\\
|\psi'_3\rangle&=&\begin{pmatrix}
\cos \frac{\chi}{2} \\
i\sin \frac{\chi}{2}\end{pmatrix},
\end{eqnarray}
with additional overall phases omitted. Here, $r$, $\theta$, and $\alpha$ are the parameters involved in the $\mathcal{PT}$-symmetric Hamiltonian given in Eq.~(\ref{Eq:Hpt}) above, and 
\begin{eqnarray}
\cos\frac{\delta}{2}&=&\frac{\cos(\omega t-\alpha)\cos\frac{\pi-2\epsilon_{12}}{4}-\sin \omega t\sin\frac{\pi-2\epsilon_{12}}{4}}{\sqrt{1-\cos(2\omega t)\sin^2\alpha+2\sin(\omega t)\sin\alpha(\cos(\omega t)\cos\alpha\sin\epsilon_{12}-\sin(\omega t)\cos\epsilon_{12})}},\\
\cos \frac{\chi}{2}&=&\frac{|\tau_1|}{|\tau_1|^2+|\tau_2|^2},\\
\tau_1&=&\cos(\omega t-\alpha)\cos\frac{\mu}{2}-i\sin \omega t e^{i\nu}\sin\frac{\mu}{2},\\
\tau_2&=& -i\sin \omega t\cos\frac{\mu}{2}+\cos(\omega t+\alpha)e^{i\nu}\sin\frac{\mu}{2}.
\end{eqnarray}
The overlaps between the normalized time-evolved states are given by
\begin{eqnarray}
|\langle|\psi'_2|\psi'_1\rangle|&=&0,\\
|\langle|\psi'_3|\psi'_1\rangle|&=&|\cos\frac{\chi+\delta}{2}|,\\
|\langle|\psi'_3|\psi'_2\rangle|&=&|\sin\frac{\chi+\delta}{2}|.
\end{eqnarray}
The orthogonality between the states $|\psi'_1\rangle$ and $|\psi'_2\rangle$ is assured by the relation between the chosen $\alpha$ in the $\mathcal{PT}$-symmetric Hamiltonian and the evolution time $t$ given by Eq.~(\ref{eqt}).

\begin{figure}[htbp!]
\centering\includegraphics[width=0.9\textwidth]{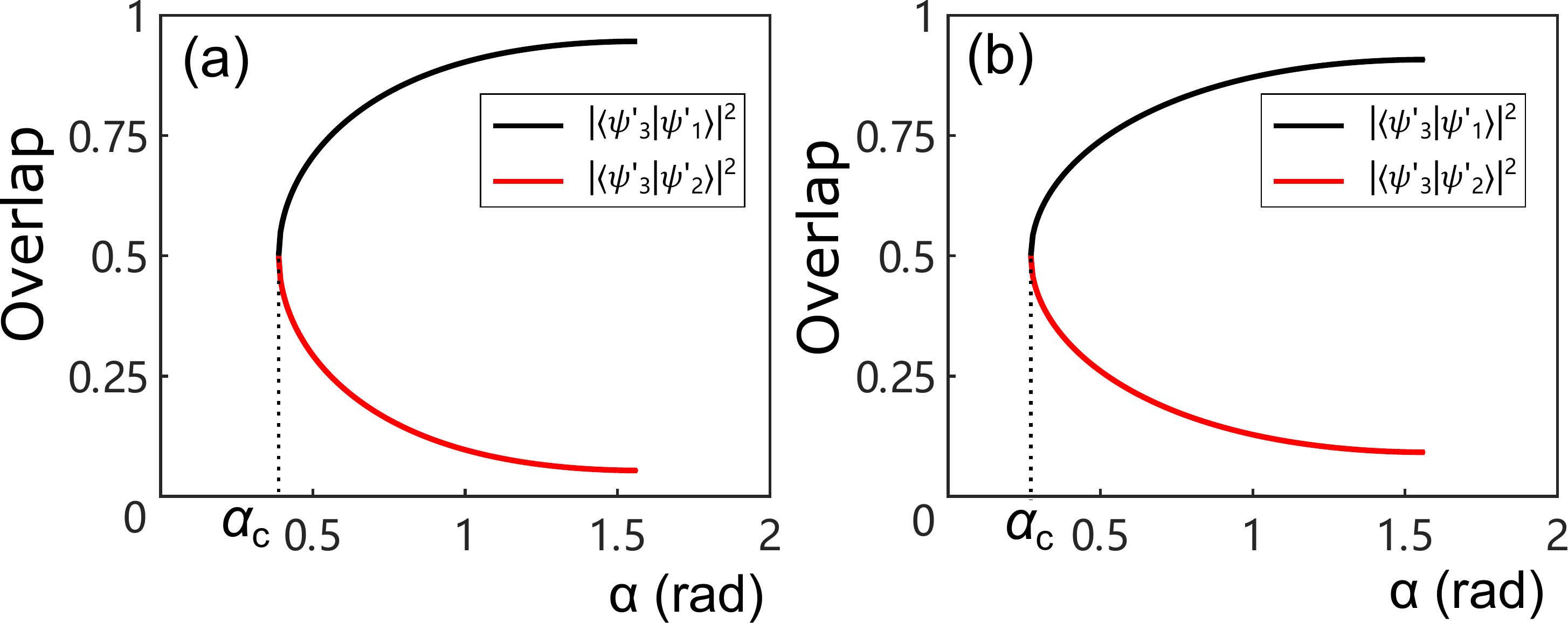}
\caption{Theoretical overlaps between the normalized time-evolved states, $|\langle|\psi'_3|\psi'_1\rangle|^2$ and $|\langle|\psi'_3|\psi'_2\rangle|^2$, with different $\alpha$ for (a) $\beta=\pi/3$ and (b) $\beta=\pi/2$ in our experiment.}
\label{figs2}
\end{figure}

Figure \ref{figs2} numerically shows the overlaps between the normalized time-evolved states, i.e., $|\langle|\psi'_3|\psi'_1\rangle|^2$ and $|\langle|\psi'_3|\psi'_2\rangle|^2$, for (a) $\beta=\pi/3$ and (b) $\beta=\pi/2$ in our experiment. One can see that as $\alpha$ tends to $\pi/2$, the overlap between the states $|\psi'_2\rangle$ and $|\psi'_3\rangle$ decreases and converges to a nonzero value.
}


\end{document}